\documentclass[10pt,journal,compsoc]{IEEEtran}
\ifCLASSOPTIONcompsoc
    \usepackage[nocompress]{cite}
\usepackage{amsmath}
\usepackage{graphicx}

\else
  \usepackage{cite}
\fi
\ifCLASSINFOpdf
\else
\fi
\begin{document}
\title{Flexible Representation and Manipulation\\of Audio Signals on Quantum Computers}
%\title{FRQA: A flexible representation of quantum audio}

\author{Fei~Yan, %~\IEEEmembership{Member,~IEEE,}
        Yiming~Guo, %~\IEEEmembership{Fellow,~OSA,}
        Abdullah~M.~Iliyasu, %~\IEEEmembership{Fellow,~OSA,}
        and~Huamin~Yang%~\IEEEmembership{Life~Fellow,~IEEE}% <-this % stops a space
\IEEEcompsocitemizethanks{\IEEEcompsocthanksitem F. Yan, Y. Guo, and H. Yang are with School of Computer Science and Technology, Changchun University of Science and Technology, No. 7089, Weixing Road, Changchun 130022, China. E-mail: yanfei@cust.edu.cn\protect\\
\IEEEcompsocthanksitem A. M. Iliyasu is with Electrical Engineering Department, College of Engineering, Prince Sattam Bin Abdulaziz University, Al-Kharj 11942, Kingdom of Saudi Arabia.}% <-this % stops an unwanted space
\thanks{}}


\markboth{}%
{Shell \MakeLowercase{\textit{et al.}}: Bare Demo of IEEEtran.cls for Computer Society Journals}



\IEEEtitleabstractindextext{%
\begin{abstract}
  By analyzing the numerical representation of amplitude values in audio signals and integrating the time component, a representation for audio signals on quantum computers, FRQA, is proposed. The FRQA representation is a normalized state that facilitates basic audio signal operations targeting these parameters. The preparation and retrieval for FRQA are discussed and, based on the FRQA state, we realize the circuits to accomplish basic audio signal operations such as signal addition, signal inversion, signal delay, and signal reversal. These operations can be employed as the major components to build advanced operations for particular applications in the quantum computing domain.
\end{abstract}

\begin{IEEEkeywords}
Quantum computing, quantum audio, audio representation, audio manipulation.
\end{IEEEkeywords}}

\maketitle

\IEEEdisplaynontitleabstractindextext

% \fi
\IEEEpeerreviewmaketitle

\IEEEraisesectionheading{\section{Introduction}\label{sec1}}

\IEEEPARstart{Q}{uantum} computation and quantum information, an area of study focused on accomplishing information processing tasks on quantum mechanical systems, has become a rapidly growing field since Feynman proposed the quantum computation model in 1982 \cite{1}. Afterwards, further developments by a number of researchers, notably, Deutsch's quantum parallelism assertion \cite{2}, Shor's integer factoring algorithm \cite{3}, and Grover's database searching algorithm \cite{4}, marked a leap forward in terms of the anticipated computing capability and secure communication for quantum computing devices. Not surprisingly, the appeal of this novel computational paradigm has continued to increase especially from scientists and engineers who work in areas related to this.

An image is a common artifact that depicts visual perception, which makes its realization and handling on the quantum computing framework an alluring pursuit. Attempts to undertake image processing based on quantum mechanics predate 2003, while those efforts had an inclination towards quantum optics and physics \cite{5,6}. Recent developments in research on quantum image processing (these with more computer science flair) developed with the proposals of quantum image representations. Since 2003, a number of quantum image representations have been proposed for encoding and transforming images on the quantum computing framework \cite{7}.

Research findings on quantum computation and quantum information have provided necessary theoretical basis for improving the possibility and reliability of quantum image processing. Nevertheless, there has not been sufficient effort to exploit another kind of information-carrying medium and its related manipulations, i.e. the quantum audio processing, to further extend the potential capabilities of quantum information processing. This is largely attributed to the lack of representations to encode quantum audio signals.

Classical audio processing may arise in either digital or analog domain, however, most modern audio systems prefer to digital representations because the techniques of digital signal processing are much more powerful and efficient \cite{8}. In order to process and transmit digital audio on the quantum computing framework, it is necessary to explore a flexible and efficient quantum representation for storing and representing digital signals. In digital signal processing domains, theoretical analyses and derivations are typically performed on discrete-time signal models created by the abstract process of sampling. Inspired by similar methods, in \cite{9}, a quantum representation of digital audio (QRDA), which uses two entangled qubit sequences to store the audio amplitude and time information, was proposed. While QRDA extends audio processing into the quantum computing domain, its numerical representation of amplitude values only offers an encoding method for unipolar (viz. non-negative) numbers which lead to operations that are susceptible to errors.

In order to develop quantum algorithms in the field of audio signal processing, a set of basic signal operations must be developed. Considering amplitude transformations required to process audio signals on quantum mechanical computers, we analyze the numerical representation of amplitude values and describe a flexible representation of quantum audio (FRQA). The FRQA encodes the amplitude information in two's complement notation and integrates the time component to a normalized quantum state, which is more flexible and better suited to be the fundamental protocol for quantum audio processing. Based on the FRQA state, audio signal operations including addition, inversion, delay, and reversal are proposed in this study. These operations are implemented by using the quantum circuit model, where algorithms are compiled into a sequence of common gates acting on one or more qubits, such as NOT, Hadamard, CNOT, and Toffoli gates \cite{10}.

The rest of the paper is organized as follows. We start with the proposal, preparation, and retrieval of the FRQA state in Section \ref{sec2}. In Section \ref{sec3}, various definitions for signal operations including addition, inversion, delay, and reversal, along with their quantum circuit constructions are introduced. Finally, concluding remarks and possible future works are discussed in Section \ref{sec4}.

\section{A flexible representation of quantum audio and its preparations}\label{sec2}
\subsection{FRQA}\label{sec2-1}
In electrical engineering, analog audio signals are often considered as a voltage that is varying over time. An Analog-to-Digital Converter (ADC) discretely takes samples from the analog signals at a given frequency (i.e. sampling rate) and according to the given binary sequence length (a.k.a. resolution), each sampled value is converted into a number based on its voltage level. In this fashion, digital audio is produced and represented as a sequence of numbers that express instantaneous amplitudes of the audio signal being sampled.

As a quantum representation, QRDA utilizes two entangled qubit sequences to store the audio amplitude and time information. Therein, the amplitude information is considered as a string of non-negative integers in the form:
\begin{equation}\label{eq1}
a_t \in \{0, 1, \ldots, 2^q-1\},
\end{equation}
where $q$ is the length of binary sequence used to store each number, $a_t$ ($t$ = 0, 1, \ldots, $L$-1) denotes the $t^{th}$ amplitude value in an $L$-sized QRDA audio.

This extension facilitates the audio representation and manipulation in quantum computing domain. Nonetheless, for reasons we adduce below, the unipolar encoding strategy used in QRDA may not be applicable for the accurate computation and processing operations in quantum audio processing:
\begin{itemize}
  \item[(1)] In QRDA, $a_t$ in Eq. (\ref{eq1}) can only represent non-negative numbers. Hence, some arithmetic operations pertaining to amplitude values are prone to errors. For instance, there are two amplitude values $a_m$ and $a_n$ ($a_m < a_n$), it is virtually impossible to implement the operation $a_m - a_n$ and depict the result.
  \item[(2)] QRDA is an unipolar representation, thus it cannot display or determine the midrange of the waveform in the processing operations. Accordingly, the single values from two waveforms are added together is by no means to give a correct mixing function, e.g., addition of the opposite amplitude values in two waveforms will accumulate to higher amplitude instead of counteracting or offsetting each other.
\end{itemize}

In this study, a flexible representation of quantum audio, i.e. FRQA, is proposed. FRQA encodes the amplitude values in quantum audio in a bipolar (both non-negative and negative) manner, i.e. $s_t \in$ \{$-2^{q-1}$, \ldots, -1, 0, 1, \ldots, $2^{q-1}-1$\}. Compared with the only known effort to perform audio signal processing on quantum computers (i.e. the QRDA), our proposal is more intuitive and graphic with the sign indicating the variation of the waveform, making it conceptually more like the real waveform of digital audio signals. More importantly, FRQA facilitates the realization of basic processing operations in an effective manner, e.g., (1) FRQA allows two sample values to be accurately added (or mixed in audio parlance) and it enables easy handling of the overflow and warp-around situations by using quantum circuits; (2) FRQA permits signal subtraction to be executed by simple logical addition operation with the inverted inputs, which offers a considerable saving in hardware complexity (since only carry logic is necessary and no borrow mechanism need be supported).

As conceived, the formalism of logic arithmetics provide FRQA with the adequate tools needed for effective quantum audio processing. Moreover, in digital audio processing, the most widespread method to represent bipolar numbers is based on using two's complement notation. Hence, the extension to encode the amplitude value in quantum audio using the two's complement arithmetic is both analogous and astute. Equation (\ref{eq2}) describes the stipulation in the form:
\begin{equation}\label{eq2}
S_t = S_t^0S_t^1\ldots S_t^{q-1}, S_t^i\in\{0, 1\}, i = 0,1,\ldots,q-1,
\end{equation}
where $t = 0, 1, \ldots, 2^l-1$ denotes the time information of a $2^l$-sized quantum audio signal, $S_t = S_t^0S_t^1\ldots S_t^{q-1}$ is the binary sequence encoding the two's complement notation of the amplitude value. Two cases of the binary sequence $S_t$ are listed as follows:
\begin{itemize}
  \item[(1)] If the amplitude value is non-negative, then $S_t^0 = 0$ and $S_t$ is simply represented as a binary sequence of the value itself.
  \item[(2)] If the amplitude value is negative, then $S_t^0 = 1$ and $S_t$ is represented by the two's complement mode of its absolute value.
\end{itemize}

Therefore, the quantum representative expression of an $2^l$-sized FRQA audio can be written as below:
\begin{equation}\label{eq3}
\vert A\rangle = {1\over 2^{l/2}}{\sum_{t=0}^{2^{l} - 1}{\vert S_t\rangle}\otimes \vert t\rangle},
\end{equation}
where $\vert S_t\rangle = \vert S_t^0S_t^1\ldots S_t^{q-1}\rangle$ is the two's complement representation of each amplitude value and $\vert t\rangle = \vert t_0t_1\ldots t_{l-1}\rangle$, $t_i\in\{0, 1\}$, is the corresponding time information. The state $\vert A\rangle$ is normalized, i.e. $\| \vert A \rangle \| = 1$. It is trivial that, as formalised in Eq. (\ref{eq3}), the FRQA needs $(q+l)$ qubits to represent a quantum audio with $2^{l}$ samples.

For an $L$-sized ($L$ cannot be represented by $2^{l}$) FRQA audio, employing $l$ qubits to represent the time information will produce $2^{l}-L$ audio redundancies \cite{9}. In this study, the amplitude values of all the redundancies are set as $\vert 0\rangle$, so the FRQA state in Eq. (\ref{eq3}) can be rewritten into the form:
\begin{equation}\label{eq4}
\vert A^\prime \rangle = {1\over 2^{l/2}}({\sum_{t=0}^{L - 1}{\vert S_t\rangle}\otimes \vert t\rangle} + {\sum_{t=L}^{2^l - 1}{\vert 0\rangle}^{\otimes q} \otimes \vert t\rangle}),
\end{equation}
where
\begin{equation}\label{eq5}
l = \left \{ \begin{array}{ll}
\lceil \log_{2}{L} \rceil , & \quad L > 1\\
1 , & \quad L = 1\\
\end{array}\right..
\end{equation}

\begin{figure}[!t]
\centering
\includegraphics[width=3.5in]{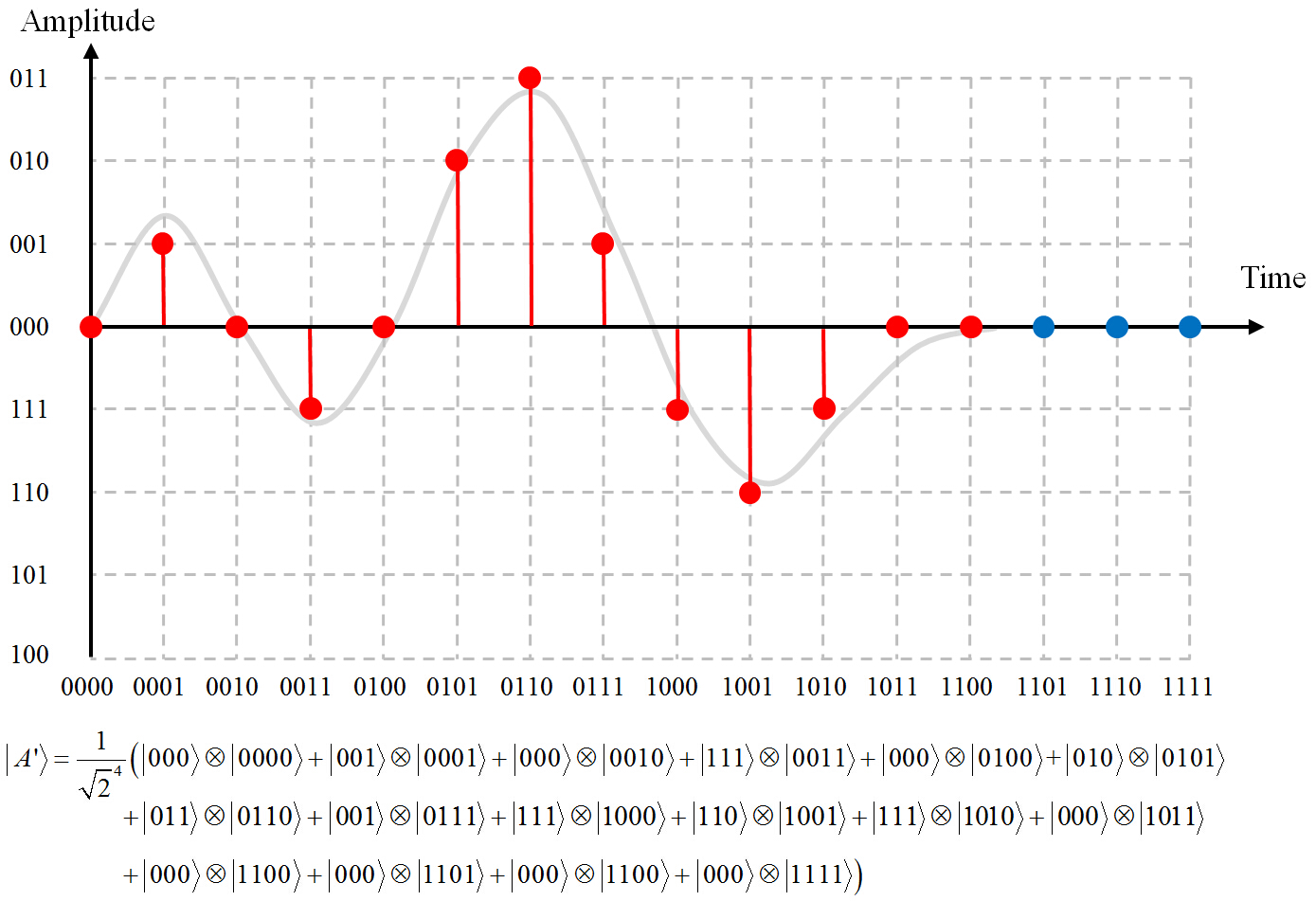}
\caption{A segment of an FRQA audio signal and its representation (blue points are the redundancies).}
\label{fig1}
\end{figure}

Figure \ref{fig1} shows a segment of an audio signal and its representative expression using the FRQA representation. In this example, amplitude values are sampled between -2 and 3, for which 3 qubits are needed to store the amplitude information in FRQA audio. The length of the audio is 13, then $l = \lceil \log_{2}{13}\rceil = 4$. Therefore, in total 7 qubits are required to represent this quantum audio signal.

\subsection{FRQA preparation}\label{sec2-2}
In quantum computation, in order to prepare a desired quantum audio signal, it is necessary to transform a quantum computer from its initialized state to the specified state. The quantum mechanical build up of quantum computers dictate that all transformations are unitary in nature. This specifies the use of unitary matrices and simple basic gates to manipulate quantum data. In this subsection, we will present a description of the procedure required to prepare an FRQA state.

As stated earlier, in digital audio, the value of each sampled point is stored on an assigned $q$-length binary sequence, i.e. $B_t^0B_t^1\ldots B_t^{q-1}$, $B_t^i\in\{0, 1\}$, which indicates the sequence is able to hold the amplitude values from 0 to $2^q-1$. In order to quantize the amplitude value ($B_t$) and convert it into the two's complement system ($\vert S_t\rangle$), we need to perform the value-setting operation $\Omega_{t}$ which could be achieved by the following two sub-operations:

\begin{itemize}
  \item[(1)] \emph{Quantization}: Using the classical binary representation of $B_t$ as the reference, the initialized states in the quantum computer could be transformed into their desired quantum states. The layout of how to accomplish this has been widely discussed in the literature, notably \cite{11,12}.
  \item[(2)] \emph{Conversion}: In order to convert the ordinary quantum state to its two's complement system, the CNOT gate operation is executed on the Most-Significant-Bit (MSB) of the qubit sequence. In this manner, the output will be the desired quantum amplitude state, i.e. $\vert S_t\rangle$.
\end{itemize}

To illustrate these steps, consider the conversion of a 3-qubit resolution to the amplitude information in quantum audio as presented in TABLE \ref{tab1}. A 3-qubit resolution holds the values ranging from 0 to 7. These numbers are binarized and quantized in two's complement notation for further computation in quantum audio processing.

\begin{table}\label{tab1}
\caption{Conversion of a 3-qubit resolution to the amplitude information in quantum audio.}
\centering
\label{tab1}
\begin{tabular}{cccc}
\hline\noalign{\normalsize}
Resolution & Binary sequence & Two's complement & Amplitude\\
\noalign{\smallskip}\hline\noalign{\smallskip}
7 & 111 & $\vert$ 011 $\rangle$ & 3 \\
6 & 110 & $\vert$ 010 $\rangle$ & 2 \\
5 & 101 & $\vert$ 001 $\rangle$ & 1 \\
4 & 100 & $\vert$ 000 $\rangle$ & 0 \\
3 & 011 & $\vert$ 111 $\rangle$ & -1 \\
2 & 010 & $\vert$ 110 $\rangle$ & -2 \\
1 & 001 & $\vert$ 101 $\rangle$ & -3 \\
0 & 000 & $\vert$ 100 $\rangle$ & -4 \\
\noalign{\smallskip}\hline
\end{tabular}
\end{table}

Therefore, the value-setting operation $\Omega_{t}$ to perform the above procedure can be executed using Eqs. (\ref{eq6}) and (\ref{eq7}):
\begin{equation}\label{eq6}
\Omega_{t} = \mathop{\otimes} \limits_{i=0}^{q-1} \Omega^{i}_{t},
\end{equation}
\begin{equation}\label{eq7}
\Omega^{i}_{t} \left( \vert 0 \rangle \right) = \left \{ \begin{array}{ll}
\vert \overline{0 \oplus B_t^i} \rangle , & \quad i = 0\\
\vert 0 \oplus B_t^i \rangle  , & \quad i \neq 0\\
\end{array}\right.,
\end{equation}
where $\oplus$ is the XOR operation. It is apparent to clarify that the operation $\Omega^{i}_{t}$ works by means of an additional CNOT gate to negate the MSB only when $B_t^i = 1$, otherwise, it remains unchanged. Following this, the quantum transformation of $\Omega_{t}$ to set amplitude value for each sample is shown in Eq. (\ref{eq8}):

\begin{equation}\label{eq8}
\begin{aligned}
%\begin{tabular}
\Omega_{t} {\vert 0 \rangle}^{\otimes q} &= \mathop{\otimes} \limits_{i=0}^{q-1} \left( {\Omega^{i}_{t}}{\vert 0 \rangle} \right) \\
&={\vert \overline{0 \oplus B_t^0} \rangle} \otimes \left(\mathop{\otimes} \limits_{i=1}^{q-1} {\vert 0 \oplus B_t^i \rangle}\right)\\
&=\mathop{\otimes} \limits_{i=0}^{q-1} {\vert S_t^i \rangle} = \vert S_t \rangle.
%\end{tabular}
\end{aligned}
\end{equation}

In this manner, a unitary transform that encodes the amplitude information by means of two's complement arithmetic is available during the preparation procedure. There are two steps to achieve the preparation procedure that are discussed as below:

\textbf{Step 1}: The initial state ${\vert 0 \rangle}^{\otimes q + l}$ is transformed into the intermediate state $\vert H \rangle$ using the 2-D identity matrix \emph{I} and 2-D Hadamard matrix \emph{H} presented in Eq. (\ref{eq9}):

\begingroup
\renewcommand\arraystretch{1.2}
\renewcommand\arraycolsep{5.0pt}
\begin{equation}\label{eq9}
I =
\left[
  \begin{array}{cc}
    1 & 0\\
    0 & 1\\
  \end{array}
\right],
H = \frac{1}{\sqrt{2}}
\left[
  \begin{array}{cc}
    1 & 1\\
    1 & -1\\
  \end{array}
\right].
\end{equation}
\endgroup

Since the tensor product of $q$ Identity matrices and $l$ Hadamard matrices are denoted by $I^{\otimes q}$ and $H^{\otimes l}$, respectively, applying the transform $\mathcal{H} = I^{\otimes q} \otimes H^{\otimes l}$ on ${\vert 0 \rangle}^{\otimes q + l}$ produces the intermediate state $\vert H \rangle$ in the form:
\begin{equation}\label{eq10}
\vert H \rangle = \mathcal{H} \left( {\vert 0 \rangle}^{\otimes q + l} \right) = \frac{1}{\sqrt{2}^{l}} {\sum_{t=0}^{2^{l}-1}} {\vert 0 \rangle}^{\otimes q} {\vert t \rangle}.
\end{equation}

So far, the time information of the FRQA model has been initialized. At this moment, the intermediate state $\vert H \rangle$ can be regarded as the superposition of all the samples of an empty digital audio, i.e. with all amplitude values set as $\vert 0\rangle$.

\textbf{Step 2}: The value-setting operation $\Omega_{t}$ is used to generate the amplitude information for each sample. Since $\Omega_{t}$ can only handle one sample at a time, considering a $2^l$-sized quantum audio, $2^l$ sub-operations are needed to execute this transformation. While $R_t$ is considered as an $l$-controlled $\Omega_{t}$ operation to integrate the amplitude values into each instant of time. For a given sample $k$, $R_t$ is defined in Eq. (\ref{eq11}):
\begin{equation}\label{eq11}
R_k = \Bigg (I \otimes {\sum_{t=0, t \neq k}^{2^l - 1}} {\vert t \rangle}{\langle t \vert} \Bigg ) + {\Omega_{k} \otimes {\vert k \rangle}{\langle k \vert}}.
\end{equation}

Applying $R_k$ on the the intermediate state $\vert H \rangle$ gives us:
\begin{equation}\label{eq12}
\begin{aligned}
R_k \left( \vert H \rangle \right) & = R_k \left( \frac{1}{\sqrt{2}^{l}} {\sum_{t=0}^{2^{l}-1}} {\vert 0 \rangle}^{\otimes q} {\vert t \rangle} \right) \\
& = \frac{1}{\sqrt{2}^{l}} R_k \left( {\sum_{t=0, t \neq k}^{2^l - 1}} {\vert 0 \rangle}^{\otimes q} {\vert t \rangle} + {\vert 0 \rangle}^{\otimes q} {\vert k \rangle} \right) \\
& = \frac{1}{\sqrt{2}^{l}} \left( {\sum_{t=0, t \neq k}^{2^l - 1}} {\vert 0 \rangle}^{\otimes q} {\vert t \rangle} + \Omega_{k} {\vert 0 \rangle}^{\otimes q} {\vert k \rangle} \right) \\
& = \frac{1}{\sqrt{2}^{l}} \left( {\sum_{t=0, t \neq k}^{2^l - 1}} {\vert 0 \rangle}^{\otimes q} {\vert t \rangle} + {\vert S_k \rangle} {\vert k \rangle} \right).
\end{aligned}
\end{equation}

From Eq. (\ref{eq12}), it is clear that for all the sub-operations $R_t$, we have:
\begin{equation}\label{eq13}
\mathcal{R} {\vert H \rangle} = \left( \prod_{t = 0}^{2^l} R_t \right) {\vert H \rangle} = {1\over 2^{l/2}}{\sum_{t=0}^{2^{l} - 1}{\vert S_t\rangle}\otimes \vert t\rangle} = \vert A \rangle.
\end{equation}

After the two steps described above, the initialized state ${\vert 0 \rangle}^{\otimes q + l}$ is transformed into the desired FRQA state by applying the unitary transform $\mathcal{R} \mathcal{H}$. Subsequently, we will continue to discuss the complexity of the preparation procedure.

\begin{figure}[!t]
\centering
\includegraphics[width=3.5in]{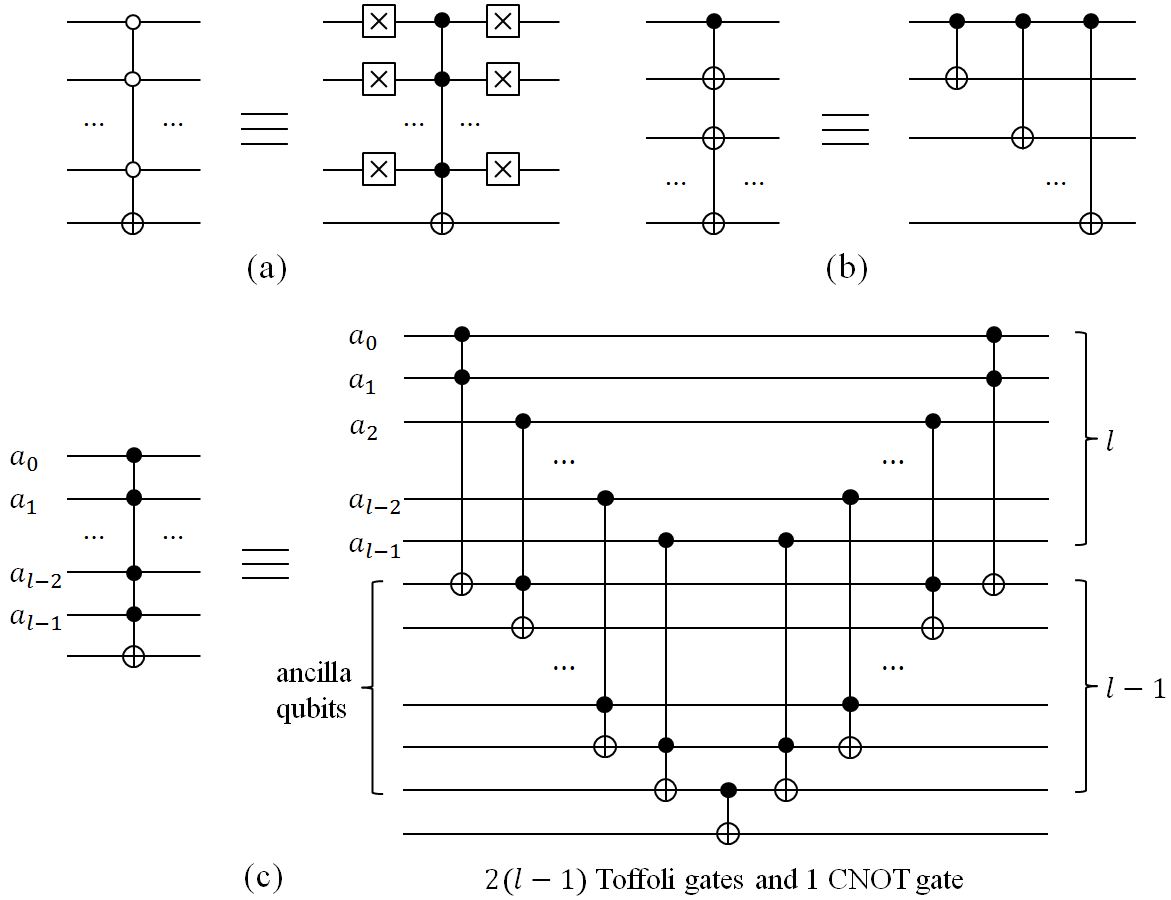}
\caption{Decomposition of some complicated quantum gates.}
\label{fig2}
\end{figure}

Complexity theory on quantum computation has been studied to discover and analyze the transformations from the basic gates, viz. the complexity of quantum algorithms is usually computed in terms of quantum gates \cite{10}. Indeed, the circuit complexity depends largely on the strategy employed for circuit decomposition and the basic operation unit that is designated to be \cite{13}. In this study, we will confine our discussion to the CNOT gate for the complexity evaluation as it is a relatively ``expensive'' elementary gate that can be easily utilized to simulate the other complicated gates. The decomposition networks for some of the complicated quantum gates used in this paper is presented in Fig. \ref{fig2}, where the $l$-controlled NOT gate can be decomposed into $2(l-1)$ Toffoli gates and 1 CNOT gate [as indicated in (c)] and the Toffoli gate can be approximately simulated by 6 CNOT gates \cite{14}.

\begin{figure}[!t]
\centering
\includegraphics[width=3.5in]{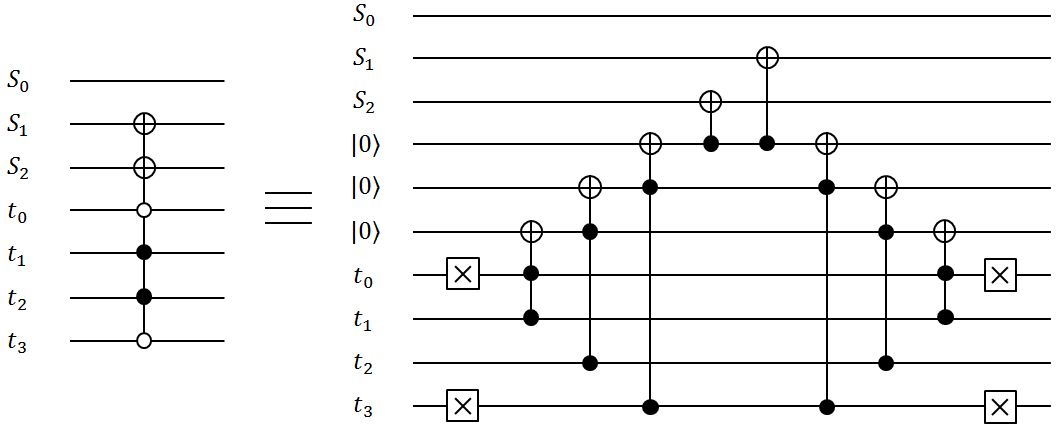}
\caption{Network of implementing the $R_6$ operation when preparing the sample at $\vert 0110\rangle$ of FRQA state in Fig. \ref{fig1}.}
\label{fig3}
\end{figure}

Based on the foregoing, it is clear that the implementation of transform $\mathcal{H}$ in Step 1 requires $l$ Hadamard gates to execute. In addition, transform $\mathcal{R}$ in Step 2 can be divided into $2^l$ sub-operations (i.e. $R_t$) to store the amplitude information for each sample. Therefore, with the assistance of enough ancillary qubits, each sub-operation $R_t$ can be directly implemented using $2\left( l - 1 \right)$ Toffoli gates and no more than $q$ CNOT gates. Hence, the complexity of preparing a $2^l$-sized FRQA state is:
\begin{equation}\label{eq14}
2^l \times \left[ 2(l - 1) \times 6 + q \right] = \left( 12l + q - 12\right) \cdot 2^l,
\end{equation}
which indicates the efficiency of the preparation process. In order to further illustrate the complexity of constructing an FRQA quantum audio, we select the initialization of the sample at $\vert 0110\rangle$ in Fig. \ref{fig1} as an example. The network for implementing the operation, i.e. $R_6$ [when \emph{k} = 6 in Eq. (\ref{eq11})], is depicted in Fig. \ref{fig3}.

\subsection{FRQA retrieval}\label{sec2-3}
The essential requirements for representing a quantum audio are the simplicity and efficiency in the storage and retrieval of the audio information. The storage of an FRQA audio is achieved by the preparation process as discussed in the preceding subsection. In this subsection, the rudiments required for the FRQA signal retrieval will be formalised.

Quantum measurement is the unique way of recovering classical information from a quantum state. To retrieve the amplitude information of each sample, quantum measurements $\Gamma$ and $M$ as defined in Eqs. (\ref{eq15}) and (\ref{eq16}) are used:
\begin{equation}\label{eq15}
\Gamma = {\sum_{t = 0}^{2^l - 1}} I^{\otimes q} \otimes {\vert t \rangle} {\langle t \vert},
\end{equation}
\begin{equation}\label{eq16}
M = {\sum_{m = 0}^{2^q - 1}} m {\vert m \rangle} {\langle m \vert}.
\end{equation}

Firstly, executing the quantum measurement operation $\Gamma$ on the time sequence extracts corresponding information of sample $t$ as $\vert P_t \rangle$, which is shown in Eq. (\ref{eq17}):
\begin{equation}\label{eq17}
\vert P_t \rangle = {\vert S_t \rangle} {\vert t \rangle},
\end{equation}
and then, the projective measurement $M$ is used to recover the amplitude value from the quantum state:
\begin{equation}\label{eq18}
\begin{aligned}
{\langle S_t \vert} M {\vert S_t \rangle} & = {\langle S_t \vert} \left( {{\sum_{m = 0}^{2^q - 1}} m {\vert m \rangle} {\langle m \vert}} \right) {\vert S_t \rangle} \\
& = {\sum_{m = 0}^{2^q - 1}} m {\langle S_t \vert} {\vert m \rangle} {\langle m \vert} {\vert S_t \rangle} = S_t.
\end{aligned}
\end{equation}

Through these measurement operations, the amplitude value of sample $k$ will be recovered. This means that all the samples in a quantum audio signal can be recovered in the same way, so the original digital audio can be retrieved from the FRQA state (Note that it is the theoretical discussion about the quantum information retrieval. We will not touch at all on the consequence of decoherence or any errors appeared in the process.)

\section{Operations on quantum audio signals}\label{sec3}
Basic signal operations are the foundations of digital audio processing. The ability to extend similar operations to quantum audio processing is essential to validating the utility of this emerging sub-field of quantum information processing. In this section, we will present a few basic audio signal operations including signal addition, signal inversion, signal delay, and signal reversal based on the FRQA representation.

\subsection{Signal addition}\label{sec3-1}
Signal addition is among the fundamental operations in audio signal processing. This operation involves the addition of amplitudes of two or more signals at each instant of time. By means of this operation, a series of audio signal processing like echo or reverb addition and active noise reduction can be implemented. Representing the amplitude values in two's complement system in FRQA facilitates determining the results of the arithmetic operations and depicting the waveform with respect to the midrange. Hence, we can define the signal addition operation based on this arithmetic advantage as below.

Assuming that $\vert A_x \rangle$ and $\vert A_y \rangle$ are two $2^l$ audio segments as presented in Eqs. (\ref{eq19})-(\ref{eq20}), the signal addition operation produces the output of $\vert A_z \rangle$ as shown in Eq. (\ref{eq21}):
\begin{equation}\label{eq19}
\vert A_x \rangle = {1\over 2^{l/2}}{\sum_{x=0}^{2^{l} - 1}{\vert S_x\rangle}\otimes \vert t_x\rangle}, \notag
\end{equation}
where
\begin{equation}
\vert t_x \rangle = \vert t_x^0t_x^1\ldots t_x^{l-1} \rangle, t_x^i\in\{0, 1\}, \notag
\end{equation}
\begin{equation}
\vert S_x \rangle = \vert S_x^0S_x^1\ldots S_x^{q-1} \rangle, S_x^i \in\{0, 1\},
\end{equation}
and
\begin{equation}\label{eq20}
\vert A_y \rangle = {1\over 2^{l/2}}{\sum_{y=0}^{2^{l} - 1}{\vert S_y\rangle}\otimes \vert t_y\rangle}, \notag
\end{equation}
where
\begin{equation}
\vert t_y \rangle = \vert t_y^0t_y^1\ldots t_y^{l-1} \rangle, t_y^i\in\{0, 1\}, \notag
\end{equation}
\begin{equation}
\vert S_y \rangle = \vert S_y^0S_y^1\ldots S_y^{q-1} \rangle, S_y^i \in\{0, 1\},
\end{equation}
so
\begin{equation}\label{eq21}
\vert A_z \rangle = {1\over 2^{l/2}}{\sum_{z=0}^{2^{l} - 1}{\vert S_z\rangle}\otimes \vert t_z\rangle}, \notag
\end{equation}
where
\begin{equation}
\vert t_z \rangle = \vert t_z^0t_z^1\ldots t_z^{l-1} \rangle, t_z^i\in\{0, 1\}, \notag
\end{equation}
\begin{equation}
\vert S_z \rangle = \vert S_z^0S_z^1\ldots S_z^{q}\rangle = \vert S_x + S_y \rangle, S_z^i \in\{0, 1\}, \notag
\end{equation}
\begin{equation}
t_z = t_x = t_y.
\end{equation}

\begin{figure}[!t]
\centering
\includegraphics[width=3in]{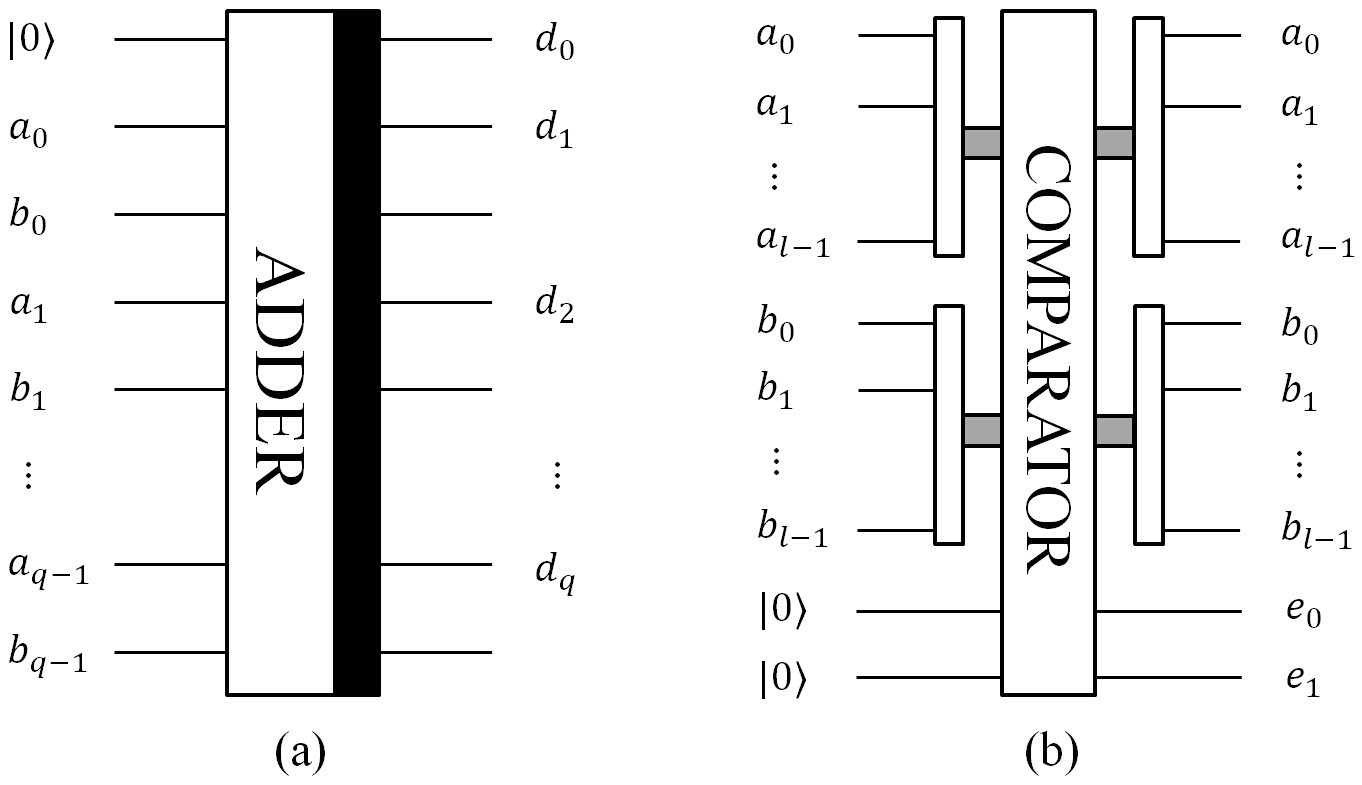}
\caption{(a) The quantum ADDER module and (b) the quantum comparator module.}
\label{fig4}
\end{figure}

Quantum algorithms are tools for constructing advanced operations using circuit elements. In order to construct the quantum circuit to execute the signal addition operation, we start by introducing two commonly-used quantum modules that are presented in Fig. \ref{fig4}.

The ADDER module \cite{13} is designed to realize the operation $\vert a \rangle\vert b \rangle \rightarrow \vert a \rangle\vert a + b \rangle$, where $\vert a \rangle$ and $\vert b \rangle$ are two quantum sequences both with $q$ qubits, and $\vert a + b \rangle$ = $\vert d_0d_1\ldots d_{q} \rangle$ is a quantum sequence with $(q + 1)$ qubits that represents the result of addition. By using the ADDER module and necessary control conditions, we can realize the addition of amplitude values in two segments of quantum audio at any given instant of time.

The other module presented in Fig. \ref{fig4} is the quantum comparator module \cite{15} that is used to compare two $l$-qubit binary sequences ($\vert a\rangle$ and $\vert b\rangle$) and stores the result in the ancilla qubits $\vert e_0e_1 \rangle$. A useful output of $\vert e_0e_1 \rangle$ in this study is when $\vert e_0e_1 \rangle = \vert 00 \rangle$, then $\vert a\rangle = \vert b\rangle$. The comparator module guarantees that the two operands in the addition are the amplitude values of the two audio segments at the same instant of time.

Hence, the operation of FRQA-based signal addition operation $\mathcal{U}_A$ can be written in the form:

\begin{equation}\label{eq22}
\mathcal{U}_A : \vert S_x \rangle\vert S_y \rangle \rightarrow \vert S_x \rangle\vert S_x + S_y \rangle.
\end{equation}

\begin{figure}[!t]
\centering
\includegraphics[width=3.2in]{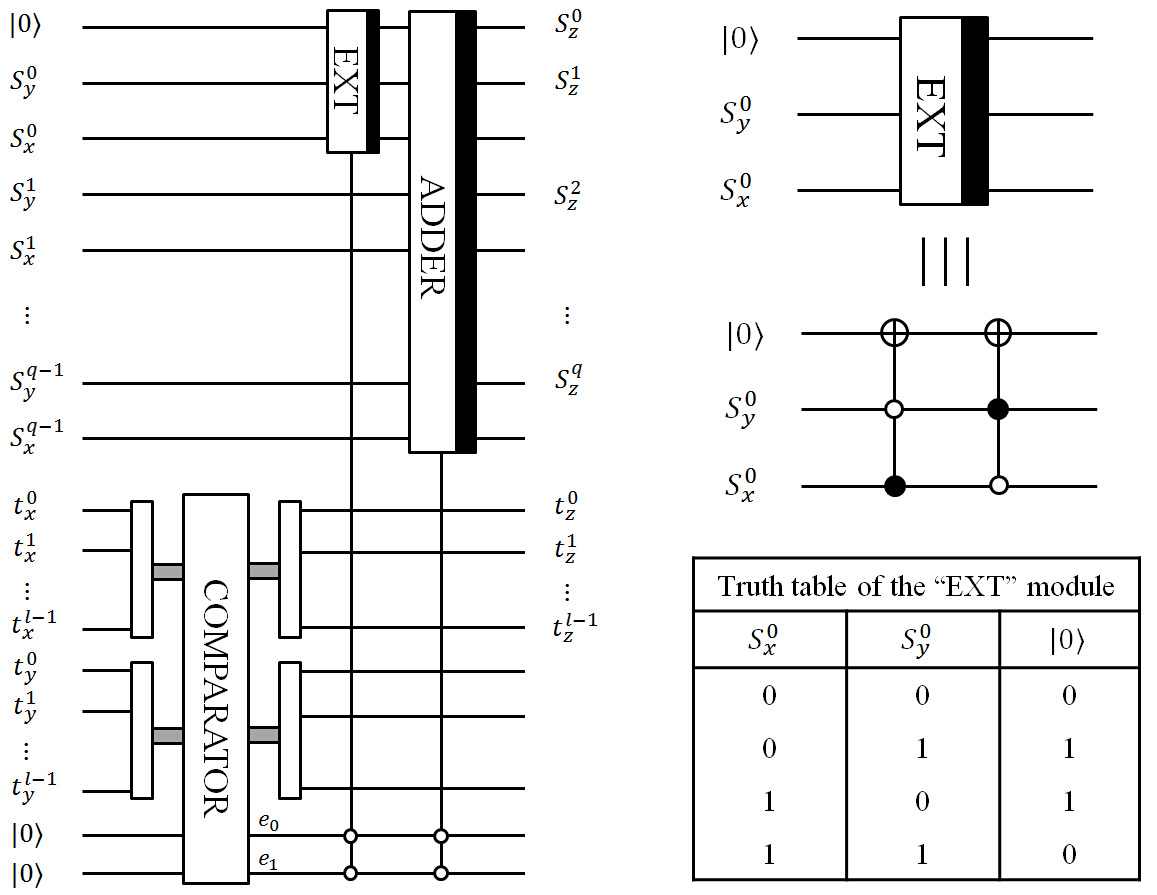}
\caption{Circuit construction to execute the quantum audio signal addition operation.}
\label{fig5}
\end{figure}

\begin{figure}[!t]
\centering
\includegraphics[width=3.5in]{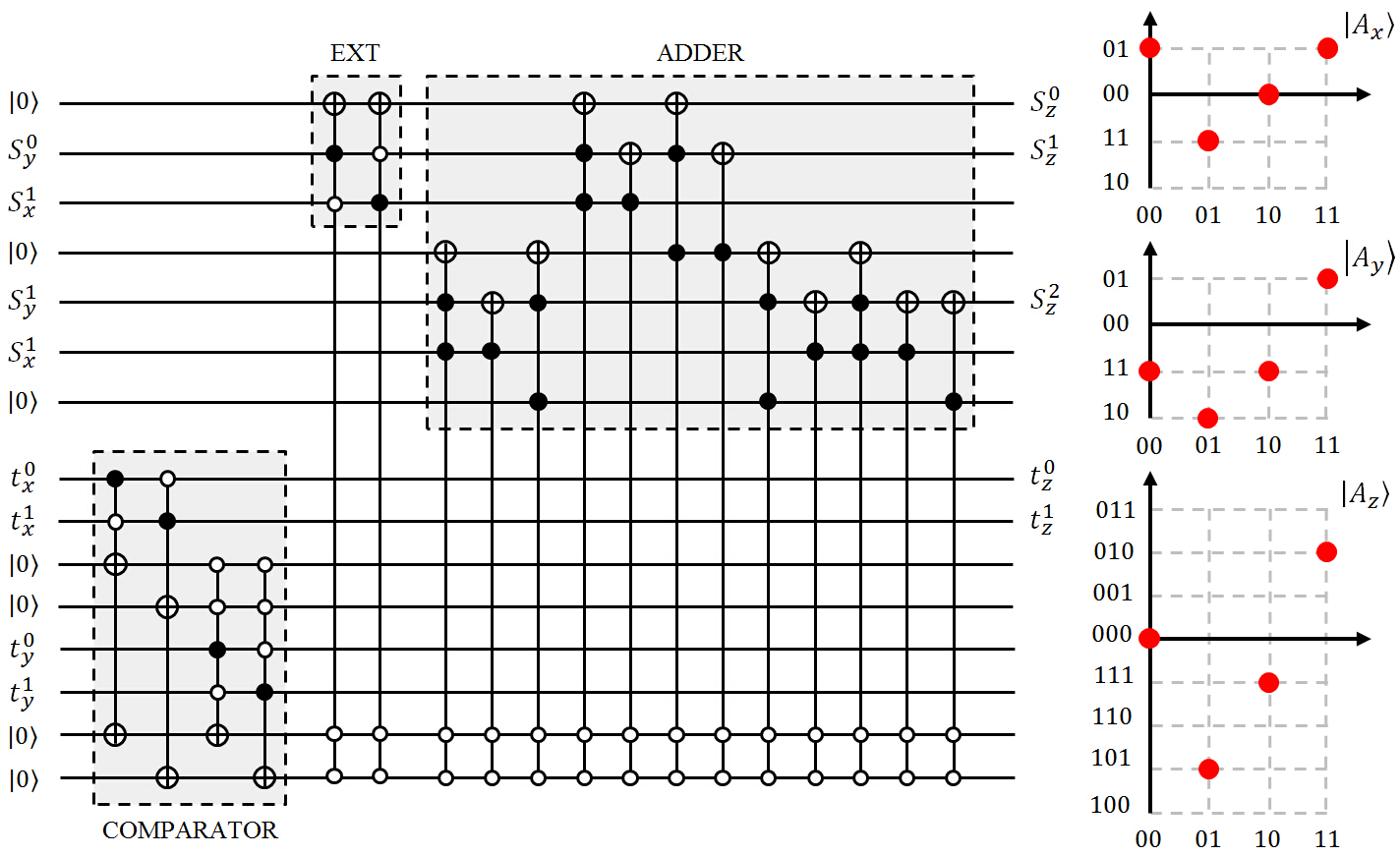}
\caption{An example of quantum audio signal addition operation.}
\label{fig6}
\end{figure}

The quantum circuit to realize the operation $\mathcal{U}_A$ is depicted in Fig. \ref{fig5}, where the ADDER operation is applied on the amplitude content and the comparator operation is applied on the time content. Additionally, in order to avoid the overflow during two's complement addition, we design a sign extension module (denoted as ``EXT" in Fig. \ref{fig5}) which helps to increase the length of the binary sequence while preserving the sign information in the ADDER module. To conclude our discussion of the signal addition operation, an example for the case when $l = 2$ and $q = 2$ is presented in Fig. \ref{fig6}.

As seen in Fig. \ref{fig5}, the circuit network for implementing the signal addition operation is divided into 3 parts: a comparator module, a $2$-controlled ``EXT" module, and a $2$-controlled ADDER module. Implementations of the ADDER and comparator module have been extensively discussed in \cite{15} and \cite{16}. Additionally, the computational complexities of the two modules are $28q - 12$ and $24l^2 + 6l$, respectively, where $q$ and $l$ denote the length of the inputs as shown in Fig. \ref{fig4}. Consequently, to effectively quantify the complexity of the operation $\mathcal{U}_A$ depicted in Fig. \ref{fig5}, we only need to determine the complexity of the $2$-controlled ``EXT" module and $2$-controlled ADDER module.

The $2$-controlled ``EXT" module is a pair of $4$-controlled NOT gates, so complexity of this part is $ 2 \times [2 \times (4-1) \times 6 + 1] = 74$. The ADDER module consists of $4q-2$ Toffoli gates and $4q$ CNOT gates \cite{16}, so the $2$-controlled unit consists of $\underline{4q-2}$ $4$-controlled NOT gates and $4q$ $3$-controlled NOT gates, thereby the complexity of this controlled module is $248q - 74$. Based on these analyses, the complexity of the entire signal addition operation is $24l^2 + 6l + 248q$.


\subsection{Signal inversion}\label{sec3-2}
Inverting a signal is common in signal processing algorithms. A meaningful usage of inversion is to transform the signal subtraction to signal addition by means of inverted inputs (audio cancellation is realized as so). In audio signal processing, signal inversion is realized by inverting all amplitude values of an audio signal. When amplitude values are represented by bipolar values, the operation is essentially an alteration of positive and negative signs. In this subsection, the FRQA-based signal inversion operation is formalised.

Assuming that $\vert A \rangle$ is an FRQA audio in the form presented in Eq. (\ref{eq3}), the signal inversion operation $\mathcal{U}_I$ applied on $\vert A \rangle$ will produce the output of $\vert A_I \rangle$ in the form shown in Eq. (\ref{eq23}):
\begin{equation}\label{eq23}
\vert A_I \rangle = {1\over 2^{l/2}}{\sum_{t=0}^{2^{l} - 1}{\vert S_I\rangle}\otimes \vert t\rangle}, \notag
\end{equation}
where
\begin{equation}
\vert t \rangle = \vert t_0t_1\ldots t_{l-1} \rangle, t_i\in\{0, 1\}, \notag
\end{equation}
\begin{equation}
\vert S_I \rangle = \vert - S_t^0S_t^1\ldots S_t^{q-1} \rangle = \vert - S_t \rangle, S_t^i \in\{0, 1\}.
\end{equation}

As in two's complement arithmetic, an effective way to negate a number is inverting all the qubits and adding ``1''. This procedure in quantum audio processing is further explained using the following steps:

\textbf{Step 1:} Invert all the qubits in $\vert S_t\rangle$ by performing operation $\mathcal{U}_{I}^1$:
\begin{equation}\label{eq24}
\begin{aligned}
\mathcal{U}_{I}^1(\vert A\rangle) & = {1\over 2^{l/2}}{\mathcal{U}_{I}^1(\sum_{t=0}^{2^{l} - 1}{\vert S_t\rangle})\otimes \vert t\rangle} \\
& = {1\over 2^{l/2}}{\sum_{t=0}^{2^{l} - 1}{\vert \overline{S_t}\rangle}\otimes \vert t\rangle}, \notag
\end{aligned}
\end{equation}
where
\begin{equation}
\vert S_t \rangle = \vert S_t^0S_t^1\ldots S_t^{q-1} \rangle, S_t^i \in\{0, 1\}, \notag
\end{equation}
\begin{equation}
\vert \overline{S_t} \rangle = \vert \overline{S_t^0 S_t^1 \ldots S_t^{q-1}} \rangle, \overline{S_t^i} \in\{0, 1\}.
\end{equation}

\textbf{Step 2:} Add ``1'' to the inverted result $\vert \overline{S_t}\rangle$ and neglect the overflow by applying operation $\mathcal{U}_{I}^2$:
\begin{equation}\label{eq25}
\mathcal{U}_{I}^2 : \vert \overline{S_t} \rangle \rightarrow = \vert S_I \rangle, \notag
\end{equation}
where
\begin{equation}
\vert \overline{S_t} \rangle = \vert \overline{S_t^0 S_t^1 \ldots S_t^{q-1}} \rangle, \notag
\end{equation}
\begin{equation}
\vert (\overline{S_t^0 S_t^1 \ldots S_t^{q-1}} + 1) \bmod 2^{q} \rangle = \vert S_I \rangle.
\end{equation}

\begin{figure}[!t]
\centering
\includegraphics[width=3in]{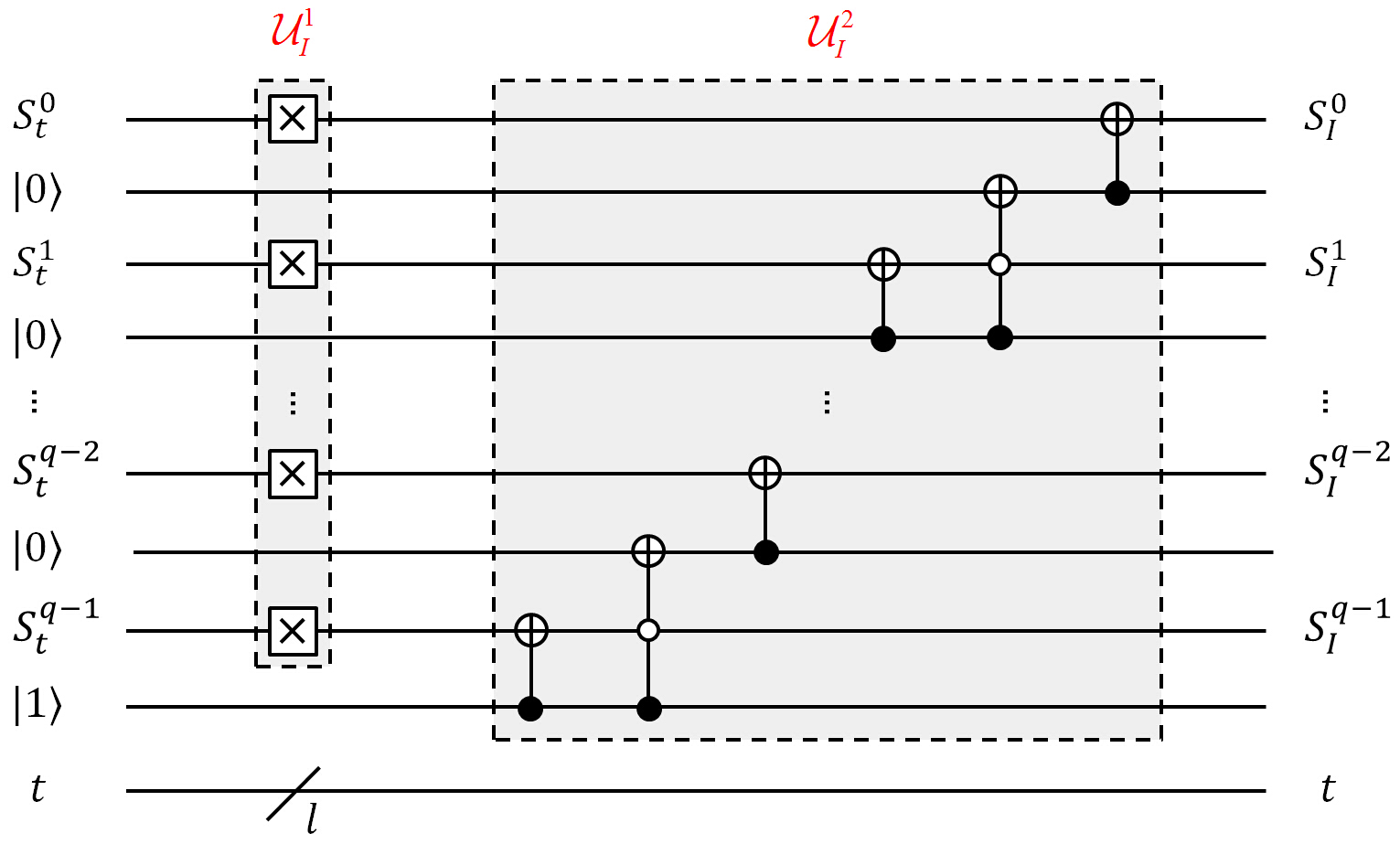}
\caption{Circuit construction to execute the quantum audio signal inversion operation.}
\label{fig7}
\end{figure}

It should be noted that similar to what is obtained in (classical) digital audio signal processing, a slight skewing of the data range is unavoidable as we cannot find the inverted counterpart of $\vert -2^{q-1}\rangle$ in the sampled amplitude values $s_t$. To overcome this, we restrict our discussion within the amplitude value range from $\vert -2^{q-1} + 1\rangle$ to $\vert 2^{q-1} - 1\rangle$. The quantum circuit for inverting an FRQA audio signal is shown in Fig. \ref{fig7}, wherein, the operation $\mathcal{U}_{I}^1$ in Step 1 can be directly implemented by using $q$ NOT gates and in the operation $\mathcal{U}_{I}^2$ in Step 2, $q - 1$ Toffoli gates and $q$ CNOT gates are required, thus the complexity of this operation is $7q - 6$.

\begin{figure}[!t]
\centering
\includegraphics[width=3.4in]{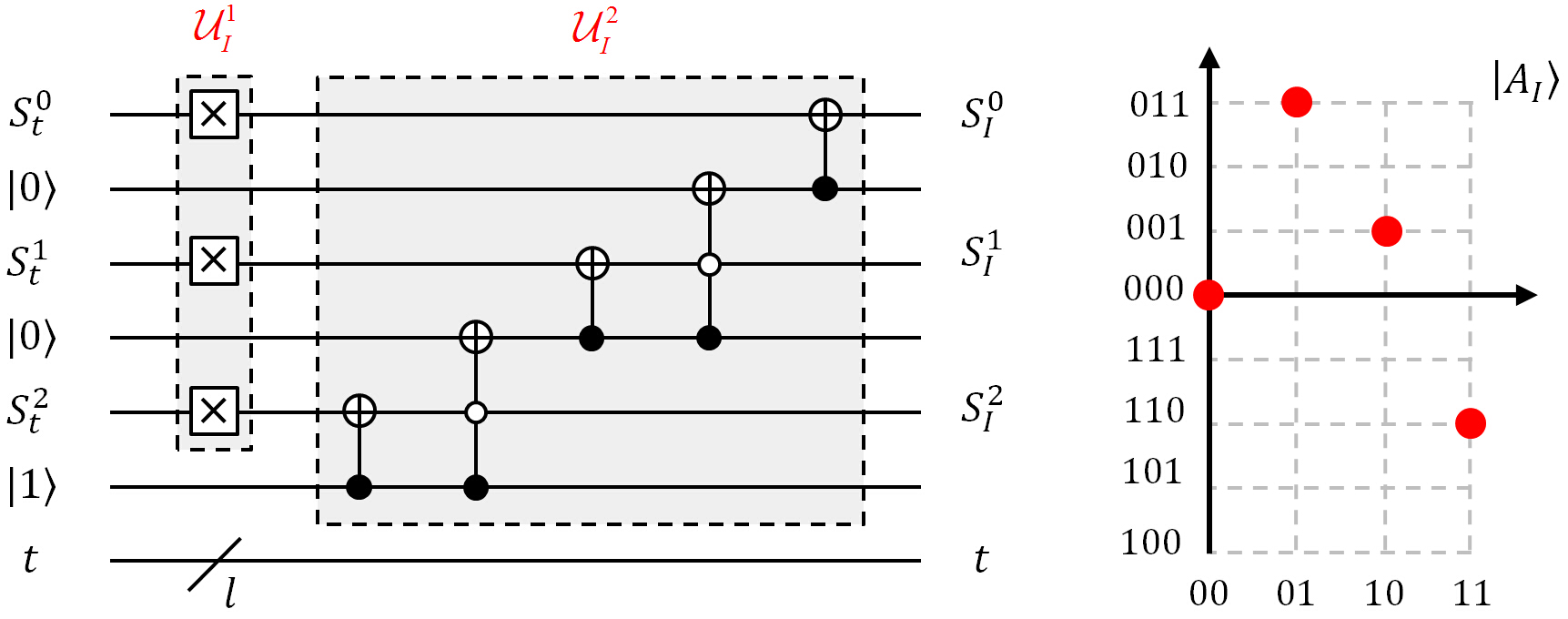}
\caption{An example of the quantum audio signal inversion operation ($\vert A_I\rangle$ is the inverted version of $\vert A_z\rangle$ in Fig. \ref{fig6}).}
\label{fig8}
\end{figure}

A simple example to illustrate how a quantum audio signal is inverted would suffice. The circuit network required to execute this procedure is presented in Fig. \ref{fig8}, where the FRQA audio $\vert A_z\rangle$ in Fig. \ref{fig6} is used as the input audio signal.

\subsection{Signal delay}\label{sec3-3}
Signal delay is the operation that records an input signal and then plays it back after a period of time. This operation is a common audio effect that is often used to create the sound of a repeating and decaying echo. In this subsection, the FRQA-based signal delay operation is formalised.

Assuming $A(t)$ is the original audio signal, then audio signal $B(t)$ is regarded as a delayed version of audio $A(t)$ if:

\begin{equation}\label{eq26}
A(t) = B(t^{\prime}), t^{\prime} = t + \triangle{t},
\end{equation}
where $t$ denotes the time information of amplitude values, $\triangle{t}$ is a fixed interval which specifies the desired delay of the system. Using the foregoing, the signal delay operation $\mathcal{U}_D$ can be described using Eqs. (\ref{eq27})-(\ref{eq28}):

\begin{equation}\label{eq27}
\mathcal{U}_{D} : A(t) \rightarrow B(t^{\prime}), t^{\prime} = t + \triangle{t},
\end{equation}
\begin{equation}\label{eq28}
B(t^{\prime}) = \left \{ \begin{array}{ll}
0 , & 0 \leq t^{\prime} \leq \triangle{t-1}\\
A(t) , & \triangle{t} \leq t^{\prime} \leq 2^{l}-1\\
\end{array}\right..
\end{equation}

Both the time information $t$ and $t^{\prime}$ of the original and delayed audio signal need to be in the range $[0, 2^l - 1]$, where apparently $\triangle{t} \leq t^{\prime} = t + \triangle{t} \leq 2^l + \triangle{t} - 1$ is beyond the interval. Therefore, we separate the time information into two parts, i.e. $[\triangle{t}, 2^l - 1]$ and $[2^l, 2^l + \triangle{t} - 1]$, for discussion.

As shown in Eq. (\ref{eq28}), when $t^{\prime} \in [\triangle{t}, 2^l - 1]$, $B(t^{\prime}) = A(t)$, in which case, the time information $t^{\prime}$ can be obtained directly by means of the ADDER module on time content. However, when $t^{\prime} \in [2^l, 2^l + \triangle{t} - 1]$, we need to employ the carry qubit of the ADDER module to achieve the interval shift $[2^l, 2^l + \triangle{t} - 1] - 2^l = [0, \triangle{t}-1]$ and set the amplitudes $S_t = 0$ as the delay period.

\begin{figure}[!t]
\centering
\includegraphics[width=3.1in]{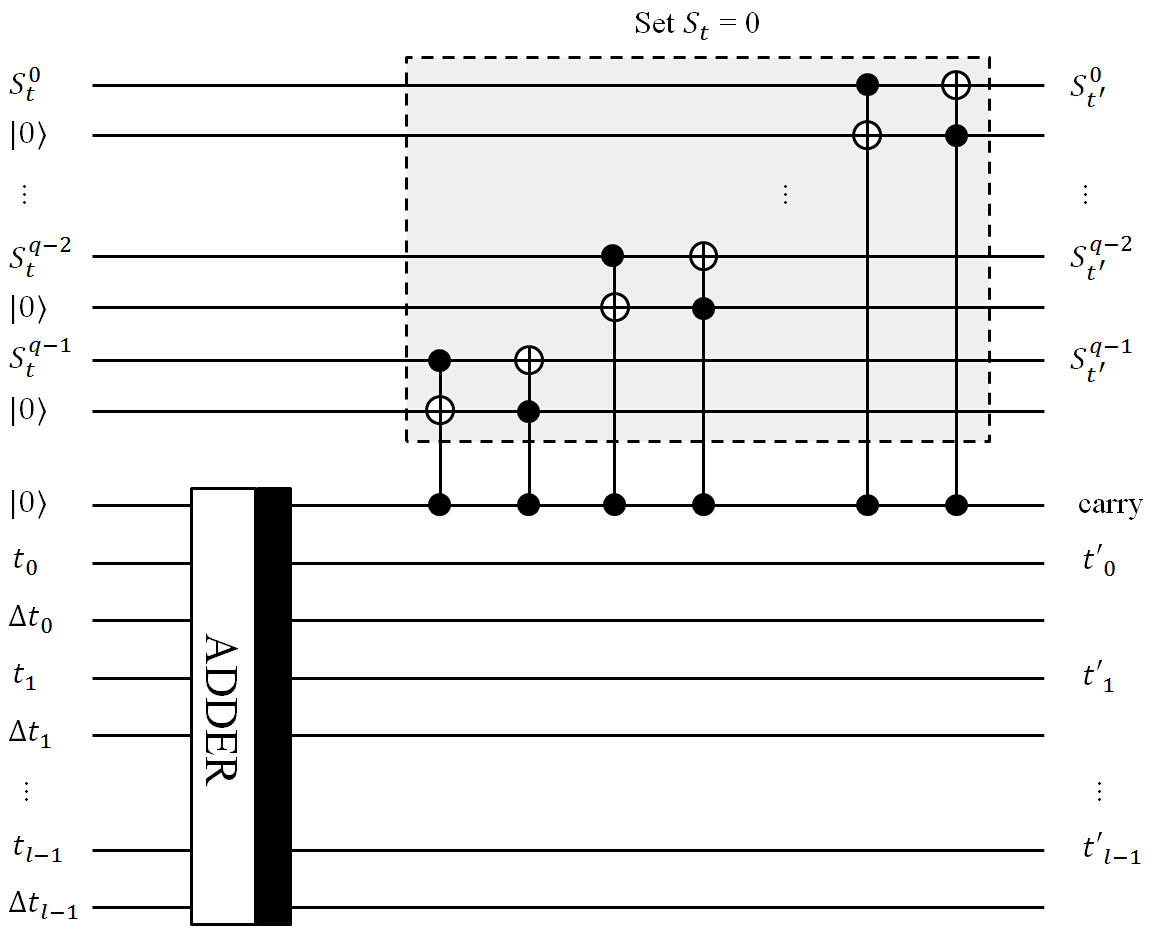}
\caption{General circuit for quantum audio signal delay operation.}
\label{fig9}
\end{figure}

\begin{figure}[!t]
\centering
\includegraphics[width=3.5in]{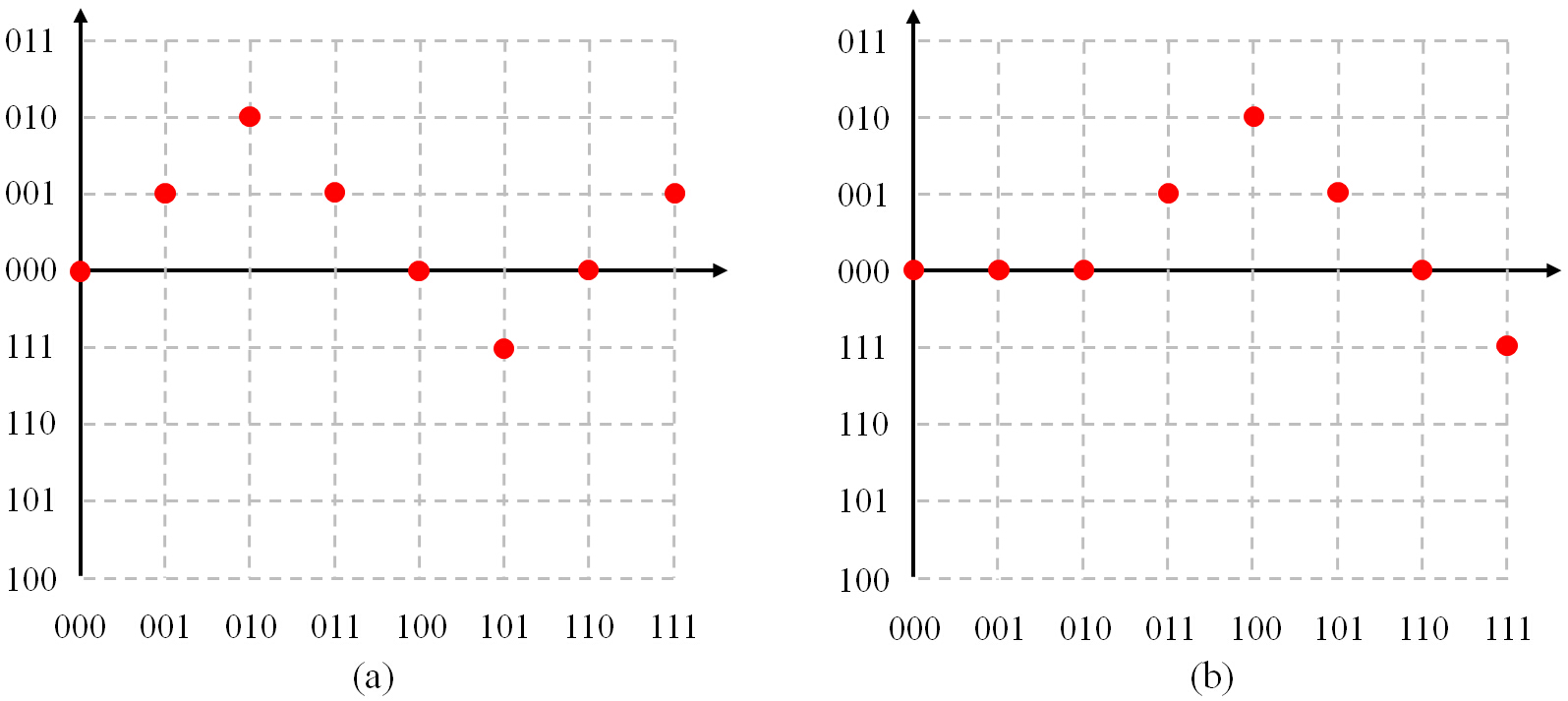}
\caption{(a) The input quantum audio signal and (b) the delayed signal.}
\label{fig10}
\end{figure}

\begin{figure}[!t]
\centering
\includegraphics[width=3.5in]{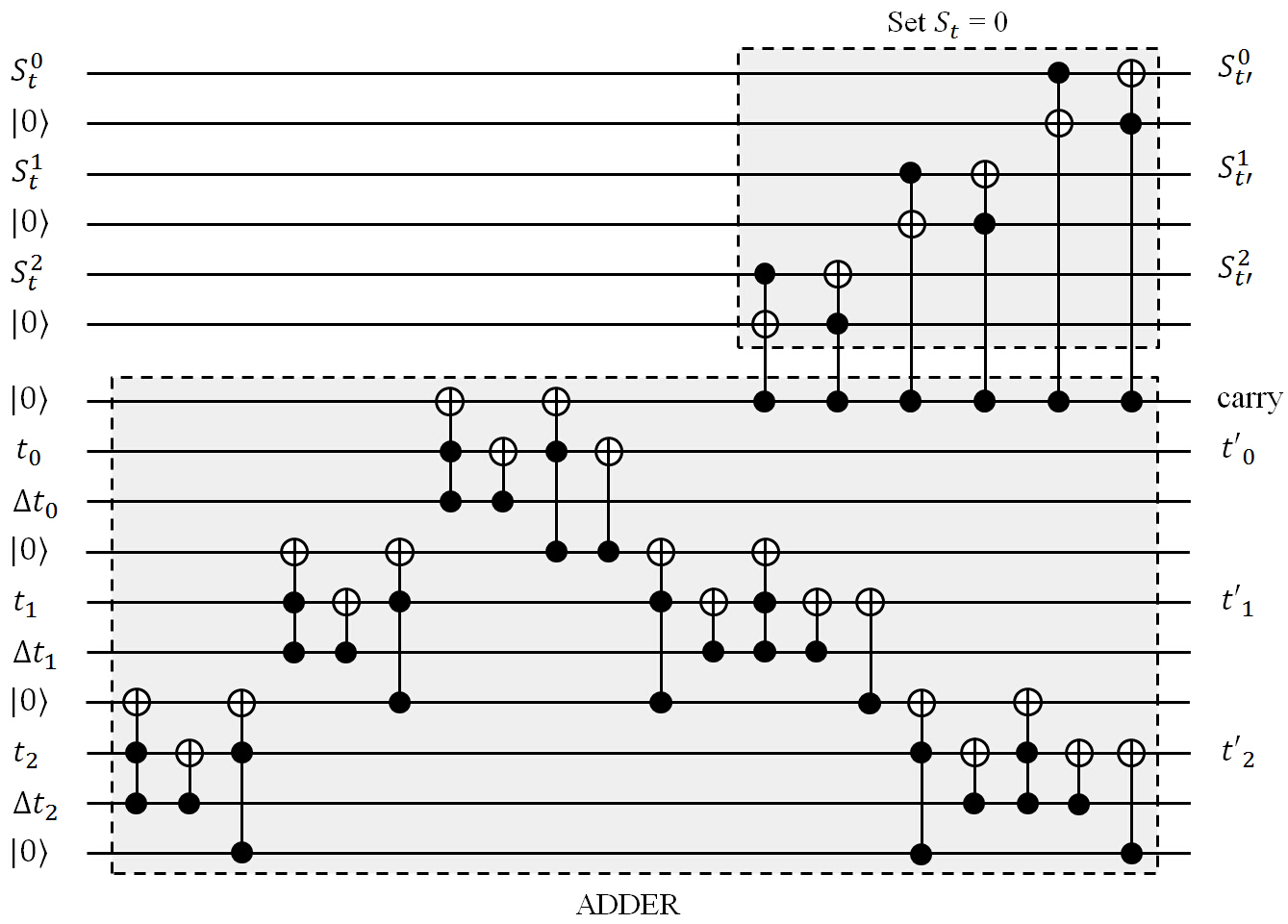}
\caption{Circuit construction to execute the quantum audio signal delay example in Fig. \ref{fig10}.}
\label{fig11}
\end{figure}

The general circuit for the quantum audio signal delay operation is presented in Fig. \ref{fig9}. In addition, Fig. \ref{fig10} presents a simple example of the FRQA-based signal delay, where (a) is the input audio signal ($l = 3$ and $q = 3$) and (b) is the signal that is delayed by 2 time units. The quantum circuit for this operation can be constructed as in Fig. \ref{fig11}.

The implementation of signal delay consists of one ADDER module (on time content) and $2q$ Toffoli gates, so the complexity of this operation is $28l + 12q - 12$.

\subsection{Signal reversal}\label{sec3-4}
Signal reversal is the process of reversing a selected audio signal such that the end of the signal is heard first and the beginning last. This operation can be used to create interesting sound effects or make small portions of inappropriate language unintelligible. Based on the FRQA state, we can define the signal reversal operation on quantum computers and construct the quantum circuit to accomplish it.

Assuming that $\vert A\rangle$ is an FRQA audio signal in the form presented in Eq. (\ref{eq3}), the signal reversal operation $\mathcal{U}_R$ applied on $\vert A \rangle$ produces an output of the form:

\begin{equation}\label{eq29}
\vert A_R \rangle = {1\over 2^{l/2}}{\sum_{t=0}^{2^{l} - 1}{\vert S_t\rangle}\otimes \vert \overline{t}\rangle}, \notag
\end{equation}
where
\begin{equation}
\vert \overline{t} \rangle = \vert \overline{t_0t_1\ldots t_{l-1}} \rangle, \overline{t_i}\in\{0, 1\}.
\end{equation}

Hence, the signal reversal operation $\mathcal{U}_R$ can be defined as below:

\begin{equation}\label{eq30}
\begin{aligned}
\mathcal{U}_R(\vert A \rangle)& = {1\over 2^{l/2}}{\sum_{t=0}^{2^{l} - 1}{\vert S_t\rangle}\otimes \mathcal{U}_R(\vert t\rangle)} \\
& = {1\over 2^{l/2}}{\sum_{t=0}^{2^{l} - 1}{\vert S_t\rangle}\otimes \vert \overline{t}\rangle}.
\end{aligned}
\end{equation}

\begin{figure}[!t]
\centering
\includegraphics[width=3.5in]{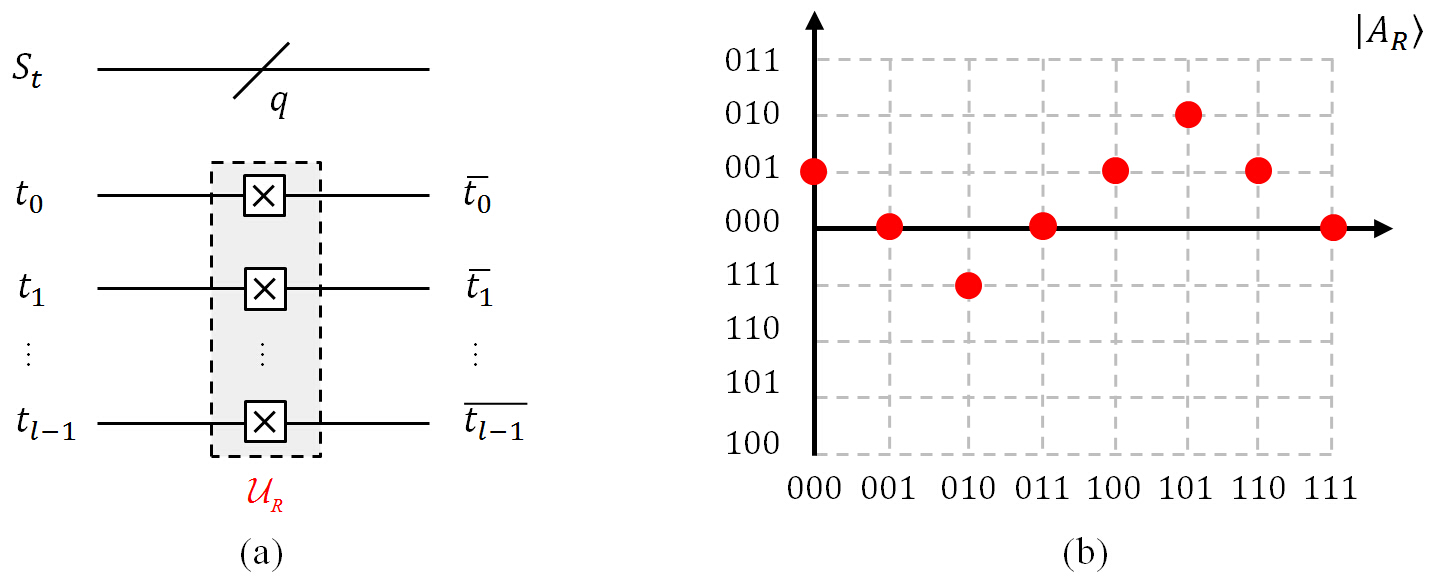}
\caption{(a) General circuit for quantum audio signal reversal and (b) a simple example of this operation.}
\label{fig12}
\end{figure}

The general circuit to execute the signal reversal operation is presented in Fig. \ref{fig12}(a). Although it is glaring that this operation can be directly implemented by using $l$ NOT gates, in order to further enunciate its realization, we use the quantum audio signal in Fig. \ref{fig10}(a) as input signal whose output is presented in Fig. \ref{fig12}(b). As seen, the result shows that all the time points are played in a reversed order.

\begin{figure}[!t]
\centering
\includegraphics[width=3.5in]{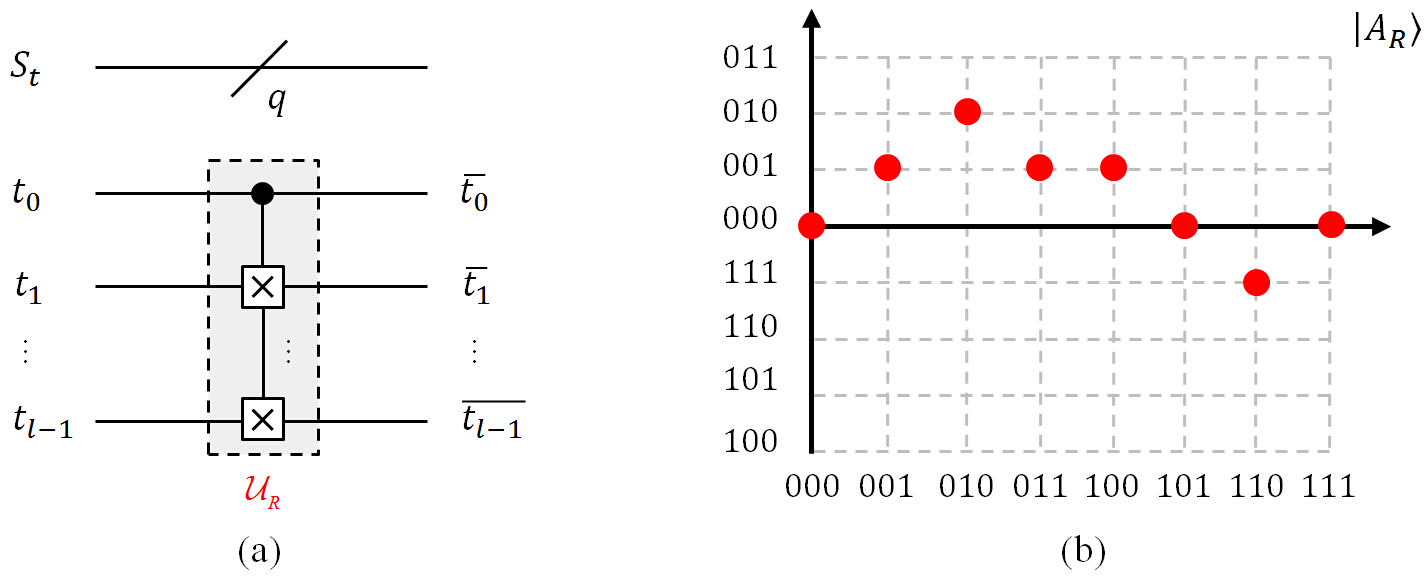}
\caption{(a) Circuit construction to execute the restricted quantum audio signal reversal and (b) a simple example of this operation.}
\label{fig13}
\end{figure}

In addition, some control condition constraints on the time content will allow us to confine the execution of the reversal operation $\mathcal{U}_R$ to a desired period. For example, in Fig. \ref{fig13}(a), the time wire $t_0$ is employed as the control wire to confine the reversal operation to the last half of the audio signal. As seen from the result in Fig. \ref{fig13}(b) that only the last 4 time points in Fig. \ref{fig10}(a) are reversed and the others are retained as in the input signal.

It should be mentioned that the complexity of the extended operation increases with the number of control qubits. In the worst case, the implementation of this operation requires an $(l-1)$-controlled NOT gate, thereby the complexity is $12l - 23$.

\section{Conclusions}\label{sec4}
The characteristics of the physical support used to embody information decide the power of a certain computational paradigm. Quantum computation has emerged as a powerful tool for efficient and secure information processing. As with any effort to process any type of information, the requirements for its encoding must be formalised. The study presented in this work explored the use of quantum mechanical properties that are at the core of quantum computers in order to encode, store, and manipulate audio signals. Inspired by the rudiments employed to capture digital audio signals, the first known effort to represent quantum audio signals (the QRDA) was broached. While QRDA extended audio processing into the quantum computing domain, its numerical representation of amplitude values only offers an encoding method for unipolar (i.e. non-negative) numbers so that its operations are susceptible to errors. The FRQA proposed in this study offers an accurate computation and extensive signal operations alternative to the QRDA. The FRQA protocol encodes amplitude values in two's complement notation so that the arithmetic advantages of doing so can be utilized to facilitate the construction of quantum circuits for amplitude transformations. By exploiting this, various operations that allow for designing advanced audio processing applications can be proposed.

Based on this new FRQA representation, basic manipulations targeting the amplitude and time components of the audio signals such as the quantum audio signal addition, inversion, reversal, and delay were formalised and insightful examples depicting their utility were also presented and discussed. The quantum circuits required to execute these basic operations on quantum audio signals were also presented and the computational requirements for their realization were analyzed.

In future work, the results from this study can be extended towards the following directions. First, considering the importance of basic operations in quantum audio processing, the operations presented in the study need to improved and new ones added to ensure the execution of more advanced algorithmic tasks. Secondly, the possibility of using quantum-based signal processing operations, such as quantum Fourier transformations \cite{14} and wavelet transformations \cite{17} which have proven well-studied in other quantum information processing applications to facilitate quantum audio manipulation needs to explored. Thirdly, the ``silent'' quantum movies proposed in \cite{18} provided impetus for the need of sound or audio on the quantum computing paradigm. A useful integration of movie strip (technically a collection of quantum images) with audio signal information will complete the quantum movie representation and facilitate more interesting applications. Finally, when the basic quantum audio signal operations are perfected, they could serve as the building block for the realization of more sophisticated algorithms used for particular applications such as watermarking, steganography, and encryption. These improvements will facilitate multimedia security during quantum information transmission.

% or

\
\ifCLASSOPTIONcompsoc
  \section*{Acknowledgments}
  This work is supported by the National Natural Science Foundation of China (No. 61502053). Additionally, AMI acknowledges funding from the Prince Sattam Bin Abdulaziz University via the Deanship for Scientific Research Project Number 2016/01/6641.

\else
  \section*{Acknowledgment}
\fi

\ifCLASSOPTIONcaptionsoff
  \newpage
\fi




\begin{thebibliography}{99}
\bibitem{1} Feynman R. Simulating physics with computers. International Journal of Theoretical Physics, 1982, 21: 467-488
\bibitem{2} Deutsch D. Quantum theory, the church-turing principle and the universal quantum computer. In: Proceedings of the Royal Society of London A, 1985, 97-117
\bibitem{3} Shor P. Algorithms for quantum computation: discrete logarithms and factoring. In: Proceedings of the 35th Annual Symposium on Foundations of Computer Science, 1994, 124-134
\bibitem{4} Grover L. A fast quantum mechanical algorithm for database search. In: Proceedings of the 28th Annual ACM Symposium on Theory of Computing, 1996, 212-219
\bibitem{5} Vlasov A. Quantum computations and images recognition. arXiv:quant-ph/9703010, 1997
\bibitem{6} Lugiato L, Gatti A, Brambilla E. Quantum imaging. Journal of Optics B: Quantum and Semiclassical Optics, 2002, 4: 176-183
\bibitem{7} Yan F, Iliyasu A, Venegas-Andraca S. A survey of quantum image representations. Quantum Information Processing, 2016, 15: 1-35
\bibitem{8} Z\"{o}lzer U. Digital audio signal processing. John Wiley \& Sons Ltd, United Kingdom, 2008
\bibitem{9} Wang J. QRDA: quantum representation of digital audio. International Journal of Theoretical Physics, 2016, 55: 1622-1641
\bibitem{10} Barenco A, Bennett C, Cleve R, et al. Elementary gates for quantum computation. Physical Review A, 1995, 52: 3457-3467
\bibitem{11} Venegas-Andraca S, Bose S. Storing, processing, and retrieving an image using quantum mechanics. In: Proceedings of SPIE Conference of Quantum Information and Computation, 2003, 134-147
\bibitem{12} Le P, Dong F, Hirota K. A flexible representation of quantum images for polynomial preparation, image compression, and processing operations. Quantum Information Processing, 2011, 10: 63-84
\bibitem{13} Vlatko V, Adriano B, Artur E. Quantum networks for elementary arithmetic operations. Physical Review A, 1996, 54: 147-153
\bibitem{14} Nielsen M, Chuang I. Quantum computation and quantum information. Cambridge University Press, United Kingdom, 2000
\bibitem{15} Wang J, Jiang N, Wang L. Quantum image translation. Quantum Information Processing, 2015, 14: 1589-1604
\bibitem{16} Jiang N, Wu W, Wang L. The quantum realization of arnold and fibonacci image scrambling. Quantum Information Processing, 2014, 13: 1223-1236
\bibitem{17} Fijany A, Williams C. Quantum wavelet transforms fast algorithms and complete circuits. Lecture Notes in Computer Science, 1999, 1509: 10-33
\bibitem{18} Iliyasu A, Le P, Dong F, et al. A framework for representing and producing movies on quantum computers. International Journal of Quantum Information, 2011, 9: 1459-1497
\end{thebibliography}
\end{document}